\documentclass[english,aps,superscriptaddress,twocolumn]{revtex4}
\usepackage[T1]{fontenc}
\usepackage[latin9]{inputenc}
\usepackage{graphicx}
\usepackage{amssymb}
\usepackage{esint}

\providecommand{\tabularnewline}{\\}

\usepackage{babel}

\begin{document}

\title{The entropy puzzle and the quark combination model}

\author{Jun Song}

\affiliation{School of Physics, Shandong University, Jinan, Shandong 250100, People's
Repulic of China}

\author{Zuo-tang Liang}

\affiliation{School of Physics, Shandong University, Jinan, Shandong 250100, People's
Repulic of China}

\author{Yu-xin Liu}

\affiliation{Department of Physics, Peking University, Beijing 100871, People's
Repulic of China}

\author{Feng-lan Shao}

\affiliation{Department of Physics, Qufu Normal University, Qufu, 273165, People's
Repulic of China}

\author{Qun Wang}

\affiliation{Interdisciplinary Center for Theoretical Study and Department of
Modern Physics, University of Science and Technology of China, Hefei
230026, China }
\begin{abstract}
We use two available methods, the Duhem-Gibbs relation and the entropy
formula in terms of particle phase space distributions, to calculate
the entropy in a quark combination model. The entropy of the system
extracted from the Duhem-Gibbs relation is found to increase in hadronization
if the average temperature of the hadronic phase is lower than that
of the quark phase. The increase of the entropy can also be confirmed
from the entropy formula if the volume of the hadronic phase is larger
than 2.5-3.0 times that of the quark phase. So whether the entropy
increases or decreases during combination depends on the temperature
before and after combination and on how much expansion the system
undergoes during combination. The current study provides an example
to shed light on the entropy issue in the quark combination model. 
\end{abstract}
\maketitle
The quark combination model \cite{Anisovich:1972pq,Bjorken:1973mh}
(QCM) or its variants such as the quark recombination or coalescence
model are used to describe multi-hadron processes in high energy electron-positron,
hadron-hadron and nucleus-nucleus collisions. Recently they are successful
in explaining the data at Relativistic Heavy Ion Collider (RHIC),
which showed that the combination is more preferable than the fragmentation
for the hadronization mechanism in intermediate transverse momentum
region in heavy ion collisions at high energies \cite{Hwa:2002tu,Greco:2003xt,Fries:2003vb}.
The basic idea of the QCM is that hadrons are combined from quarks
and anti-quarks, i.e. two quarks merge into a meson (a 2-to-1 process)
and three quarks into a baryon (a 3-to-1 process). One of the most
frequently asked questions about the validity of the QCM is: does
the entropy decrease in such 2-to-1 and 3-to-1 processes? The question
arises naturally from the common sense that the entropy is normally
reduced in a particle number decreasing process. The entropy problem
is essential to nucleus-nucleus collisions since the issue of the
entropy production is closely related to thermalization and has been
the concern of the researchers for many years. There have been previous
attempts or arguments about the entropy in the QCM-like models in
literature \cite{Greco:2003mm,Nonaka:2005vr,Hwa:2004ng,Hwa:2009bh},
but no systematic investigation has yet been made. In this paper we
will address the entropy problem in the course of the quark combination
in heavy ion collisions. As an example we will use a specific version
of the QCM which is developed by the Shandong group \cite{Xie:1988wi,Liang:1991ya,Wang:1995ch,Wang:1996pg}.
The advantage of the Shandong QCM is its exclusive nature and simplicity,
i.e. it can give all hadrons in an event. Considering that the entropy
can be regarded as a measure of degrees of freedom, the exclusive
nature of Shandong QCM is suitable and applicable to the entropy problem. 

The building blocks for the QCM are constituent quarks and anti-quarks
of the light flavors $u$,$d$ and $s$ with the momentum distributions
$f_{i}(\mathbf{p})\equiv f_{i}(y,\mathbf{p}_{T})$ \cite{Shao:2004cn,Song:2007cx,Shao:2009uk}.
All hadrons come out with fixed momenta from combination of their
constituent quarks. Note that our QCM (or any of QCM-like models)
does not deal with how the constituent quarks acquire masses which
is closely related to the physics of chiral phase transition, although
the QCM-like models can be regarded as a phenomenological and microscopic
description of the phase transition. There are no gluons in the model
because the role of a gluon can be replaced by a pair of quarks. Actually
hadronization is a dynamic process which takes place locally at a
freeze-out temperature, while our QCM (or any of QCM-like models)
lacks the real time feature but treats hadronization in a global or
an effective way. In this sense the QCM is similar to the statistical
models. 

Suppose we have $N_{i}$ quarks or anti-quarks with the flavor $i=u,d,s,\overline{u},\overline{d},\overline{s}$,
the total number of quarks and anti-quarks are then $N_{q\overline{q}}=N_{q}+N_{\overline{q}}$
with $N_{q}=N_{u}+N_{d}+N_{s}$ and $N_{\overline{q}}=N_{\overline{u}}+N_{\overline{d}}+N_{\overline{s}}$.
In mid-rapidity and high energies almost all quarks and anti-quarks
are excited from vacuum in pairs, we have $N_{q}\approx N_{\overline{q}}$.
In lower energies there are remenants of nucleons participating in
the collisions which are left in the mid-rapidity region and leads
to $N_{q}>N_{\overline{q}}$. Since the strange quarks are heavier
they are less produced and we can use a factor $\lambda_{s}=N_{s}/N_{u}\approx N_{s}/N_{d}<1$
to characterize the suppression of strange quarks relative to up and
down quarks. The procedure to randomly combine all quarks and anti-quarks
into hadrons is quite straightforward: we line up all these quarks
and anti-quarks randomly into a 1-dimensional queue and let them combine
into the groud state hadrons following the combination rule \cite{Xie:1988wi,Liang:1991ya,Wang:1995ch,Wang:1996pg}.
The momentum distribution of a specific hadron is the convolution
of those of its constituent quarks. Here the groud state hadrons include
$(3/2)^{+}$ decuplet and $(1/2)^{+}$ octet baryons and $0^{-}$
pseudoscalar and $1^{-}$ vector mesons. The short-life resonances
are allowed to decay into the long-life or final state hadrons which
are recorded in the detectors. The experimental data are used to fix
the inputs of the model. The charged multiplicity can fix $N_{q}$
and $N_{\overline{q}}$. The ratio of multiplicity of kaons to pions
can fix $\lambda_{s}$. Note that we focus on the unit rapidity region
at the central rapidity $|y|<0.5$. One of the most important element
of the model is the transverse momentum spectra of thermal quarks
which are parameterized by the blast wave model \cite{Schnedermann:1993ws},
\begin{eqnarray}
f_{i}(y,\mathbf{p}_{T}) & \propto & \int_{0}^{1}d\xi\xi\sqrt{p_{T}^{2}+m_{i}^{2}}I_{0}\left(\frac{p_{T}\sinh\rho}{T_{f}}\right)\nonumber \\
 &  & \times K_{1}\left(\frac{p_{T}\cosh\rho}{T_{f}}\right),\end{eqnarray}
where $\xi=r/R_{max}$ with $R_{max}$ the maximum radius of the thermal
source, $T_{f}=165$ MeV is the local temperature, $\rho=\tanh^{-1}(\beta_{s}\xi^{n})$
with $\beta_{s}$ the radial velocity on the surface of the firsball
and $n=0.3$ a parameter, $I_{0}$ and $K_{1}$ are the modified Bessel
functions. The average radial velocity is given by $\left\langle \beta_{r}\right\rangle =2\beta_{s}/(n+2)$.
By fitting the data for transverse momentum spectra of hadrons we
can fix the parameters $\beta_{s}$ or equivalently $\left\langle \beta_{r}\right\rangle $,
see Tab. \ref{tab:parameters}. With these inputs the available data
for the transverse momentum spectra and rapidity density of hadrons
have been well described \cite{Shao:2004cn,Song:2007cx,Shao:2009uk}.
The reason we only consider thermal quarks is that the entropy is
a bulk property which is dominated by the small $p_{T}$ region. 

Equipped by the above settings, we are able to calculate the entropy
of the system before and after hadronization. Given all particle momentum
distributions, one way to obtain the entropy of a system is through
the Duhem-Gibbs relation \cite{Demir:2008tr}, \begin{equation}
S=\frac{1}{T}(E+PV-\sum_{i}\mu_{i}N_{i}),\label{eq:entropy}\end{equation}
where $S,$ $E$, $P$, $V$ and $T$ are the entropy, energy, pressure,
volume, and temperature of the system respectively, and $\mu_{i}$
and $N_{i}$ are the chemical potential and number for the quark/hadron
species $i$ respectively. Note that the system will undergo expansion
during hadronization which is indeed a dynamic and real time process
and takes place in local volume. The global volume of an expanding
system is not well defined in a Lorentz invariant way. Here the volume
$V$ is only an effective parameter to characterize the average particle
occupancy in spatial dimension. The chemical potential $\mu_{i}$
can be determined from the multiplicity ratio of the particle $i$
and to its anti-particle $\overline{i}$ via $N_{i}/N_{\overline{i}}=e^{2\mu_{i}/T}$
where $T$ is set to 170 MeV for all particles. The quantities on
the right hand side of Eq. (\ref{eq:entropy}) are all known with,
\begin{eqnarray}
E & = & \sum_{i}\int d^{3}p\varepsilon_{i}f_{i}(\mathbf{p}),\nonumber \\
PV & = & \sum_{i}\int d^{3}p\frac{p^{2}}{3\varepsilon_{i}}f_{i}(\mathbf{p}),\nonumber \\
N_{i} & = & \int d^{3}pf_{i}(\mathbf{p}),\label{eq:e-pv-n}\end{eqnarray}
where $p\equiv|\mathbf{p}|$ and $\varepsilon_{i}=\sqrt{p^{2}+m_{i}^{2}}$
denotes the scalar momentum and energy respectively. For constituent
quarks we choose $m_{u}=m_{d}=330$ MeV and $m_{s}=550$ MeV. 

In Tab. \ref{tab:entropy-per-y} we list the total $E$, $PV$, $\mu N\equiv\sum_{i}\mu_{i}N_{i}$
and $TS$ per unit rapidity in the central rapidity region $|y|<1$
for the quark and hadronic phases. We can see that $(PV)_{h}$ (for
hadrons) is slightly smaller than $(PV)_{q}$ (for quarks) because
\begin{eqnarray}
 &  & \frac{p_{1}^{2}}{3\varepsilon_{1}}+\frac{p_{2}^{2}}{3\varepsilon_{2}}-\frac{(p_{1}+p_{2})^{2}}{3(\varepsilon_{1}+\varepsilon_{2})}\nonumber \\
 & = & \left[(p_{1}\varepsilon_{2}-p_{2}\varepsilon_{1})^{2}+2p_{1}p_{2}\varepsilon_{1}\varepsilon_{2}\left(1-\frac{\mathbf{p}_{1}\cdot\mathbf{p}_{2}}{p_{1}p_{2}}\right)\right]\nonumber \\
 &  & /[3\varepsilon_{1}\varepsilon_{2}(\varepsilon_{1}+\varepsilon_{2})]\nonumber \\
 & \approx & \frac{(p_{1}\varepsilon_{2}-p_{2}\varepsilon_{1})^{2}}{3\varepsilon_{1}\varepsilon_{2}(\varepsilon_{1}+\varepsilon_{2})}\gtrsim0,\end{eqnarray}
where in the last line is due to that quarks with the same momentum
direction combine. We also see that $\mu N$ is very small compared
to $PV$ and $E$ since in high energy collisions almost all quarks
in central rapidity are excited in vacuum. Note that the energy per
unit rapidity are not exactly equal for the two phases and the difference
is within a few percent. The effect of the energy non-conservation
is the result of the momentum conservation imposed during the course
of a specific combination, because energy and mometum conservation
cannot be simultaneously fullfilled with fixed quark masses. For example
two light constituent quarks with the total mass 660 MeV combine into
a pion of mass 140 MeV, the energy of pion is less than that of two
quarks for small momenta. But the non-conservation effect is small
on average. With $(PV)_{q}\approx(PV)_{h}$ and $V_{q}<V_{h}$ we
obtain $P_{q}>P_{h}$ and then $P_{q}-B=P_{h}$ ($B$ is the bag constant),
which is consistent to the condition for phase transition. Finally
we see that $TS$ per unit rapidity is almost constant for two phases
at each collisional energy upto a few percent. If we assume a sudden
hadronization the temperature should be the same for the two phases,
then we have the approimate conservation of the total entropy in hadronization.
Actually the hadronization time is finite during which the system
expands and cools down. Therefore the total entropy $S$ of the hadronic
phase (with lower temperature) is larger than that of the quark phase
(with higher temperature). All above observations are about the directly
produced hadrons right after the combination. For the final state
hadrons after resonance decays we also see that $E$, $PV$, $\mu N$
and $TS$ do not change much as compared to the directly produced
hadrons. This observation that the resonance decays cannot compensate
much to the total entropy is different from previous arguments that
resonant decays might be important. 

\begin{table}
\caption{\label{tab:entropy-per-y}Total $E$, $PV$, $\mu N\equiv\sum_{i}\mu_{i}N_{i}$
and $TS$ per unit rapidity in the central rapidity region $|y|<1$
for the quark and hadronic phases. The unit is GeV. The direct hadrons
are those directly from combination without any resonance decays.
The final hadrons are long-life hadrons including the contributions
from resonance decays, namely $\pi^{\pm,0}$, $K^{\pm}$, $K^{0}$,
$\overline{K}^{0}$, $p$, $\overline{p}$, $n$ and $\overline{n}$. }

\begin{tabular}{|c|c|c|c|c|c|c|c|c|}
\hline 
 & \multicolumn{4}{c|}{quark} & \multicolumn{4}{c|}{direct hadron}\tabularnewline
\hline
\hline 
$\sqrt{s_{NN}}$ & $E$ & $PV$ & $N\mu$ & $TS$ & $E$ & $PV$ & $N\mu$ & $TS$\tabularnewline
\hline 
17.3 & 459.4 & 112.9 & 10.8 & 561.5 & 456.5 & 91.5 & 15.4 & 532.6\tabularnewline
\hline 
62.4 & 609.6 & 155.5 & 1.48 & 763.7 & 615.0 & 132.4 & 2.0 & 745.5\tabularnewline
\hline 
130 & 873.6 & 226.1 & 0.34 & 1099.4 & 861.0 & 191.5 & 0.46 & 1052.0\tabularnewline
\hline 
200 & 939.7 & 248.1 & 0.21 & 1187.6 & 951.0 & 216.5 & 0.29 & 1167.2\tabularnewline
\hline
\end{tabular}

\begin{tabular}{|c|c|c|c|c|}
\hline 
 & \multicolumn{4}{c|}{final hadron}\tabularnewline
\hline
\hline 
$\sqrt{s_{NN}}$ & $E$ & $PV$ & $N\mu$ & $TS$\tabularnewline
\hline 
17.3 & 424.7 & 110.2 & 18.7 & 516.2\tabularnewline
\hline 
62.4 & 564.4 & 152.5 & 2.4 & 714.5\tabularnewline
\hline 
130 & 732.5 & 199.7 & 0.6 & 931.6\tabularnewline
\hline 
200 & 884.1 & 244.6 & 0.37 & 1128.3\tabularnewline
\hline
\end{tabular}
\end{table}

Another way of obtaining the entropy is directly through the formula,
\begin{eqnarray}
S & = & \sum_{i}d_{i}\int\frac{d^{3}rd^{3}p}{(2\pi)^{3}}\left\{ -g_{i}(\mathbf{r},\mathbf{p})\ln g_{i}(\mathbf{r},\mathbf{p})\right.\nonumber \\
 &  & \left.\pm[1\pm g_{i}(\mathbf{r},\mathbf{p})]\ln[1\pm g_{i}(\mathbf{r},\mathbf{p})]\right\} ,\label{eq:fermi-dirac}\end{eqnarray}
where $d_{i}$ are degeneracy factors for quark or hadron species
$i$, the signs $+/-$ are for bosons/fermions, and $g_{i}(\mathbf{r},\mathbf{p})$
are phase-space distributions satisfying $\int\frac{d^{3}r}{(2\pi)^{3}}d_{i}g_{i}(\mathbf{r},\mathbf{p})=f_{i}(\mathbf{p})$.
In the classical limit where $g_{i}(\mathbf{r},\mathbf{p})\ll1$,
one reaches, \begin{equation}
S\approx\sum_{i}d_{i}\int\frac{d^{3}rd^{3}p}{(2\pi)^{3}}\left[-g_{i}(\mathbf{r},\mathbf{p})\ln g_{i}(\mathbf{r},\mathbf{p})+g_{i}(\mathbf{r},\mathbf{p})\right].\label{eq:classical}\end{equation}
We can take pions which are most populated particles as an example
to estimate the order of magnitude of the phase space distribution.
The multiplicity rapidity density of $\pi^{+}$ in most central collisions
(centrality 0-5\%) at 200 GeV A is about $320$. Suppose most pions
are limited inside a momentum volume of the size $(\Delta p)^{3}\sim2\pi p_{T}^{2}\cosh y(\Delta p_{T}\Delta y)\sim10\;\mathrm{GeV}^{3}$
and inside the spatial volume of about $V_{h}\sim2200\:\mathrm{fm}^{3}$,
then we get $g_{\pi}(\mathbf{r},\mathbf{p})\sim0.027$. This provides
an upper limit, for other particles $g_{i}(\mathbf{r},\mathbf{p})\ll1$,
therefore Eq. (\ref{eq:classical}) is a quite good approximation
of Eq. (\ref{eq:fermi-dirac}). 

\begin{table}
\caption{\label{tab:parameters}Radial flow parameters in the transverse momentum
spectra for quarks at four collisional energies. The unit for energies
is GeV. }

\begin{tabular}{|c|c|c|}
\hline 
$\sqrt{s_{NN}}$ & $\left\langle \beta_{r}\right\rangle $ for u,d & $\left\langle \beta_{r}\right\rangle $ for s\tabularnewline
\hline
\hline 
17.3 & 0.38 & 0.416\tabularnewline
\hline 
62.4 & 0.483 & 0.5\tabularnewline
\hline 
130 & 0.537 & 0.55\tabularnewline
\hline 
200 & 0.562 & 0.587\tabularnewline
\hline
\end{tabular}
\end{table}

\begin{figure}
\caption{\label{fig:entropy-ratio}(Color online) The entropy ratio $R=S_{h}/S_{q}$
as functions of the volume ratio $x=V_{h}/V_{q}$ at four collisional
energies. }

\includegraphics[scale=0.37]{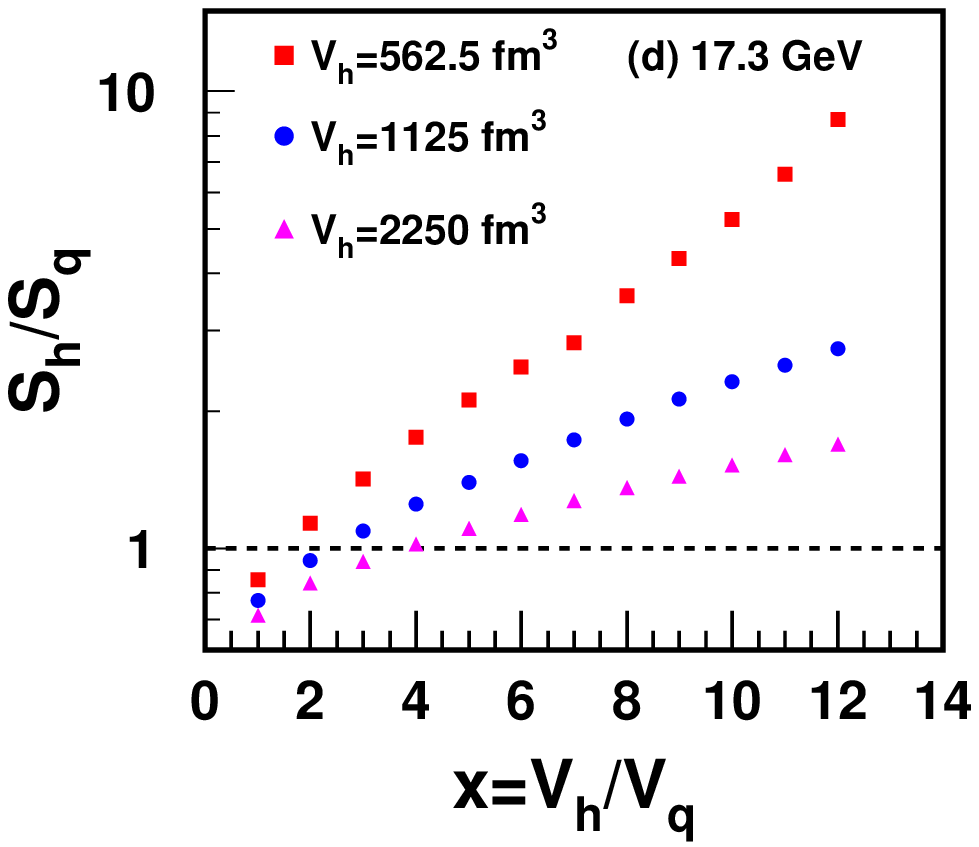}\includegraphics[scale=0.37]{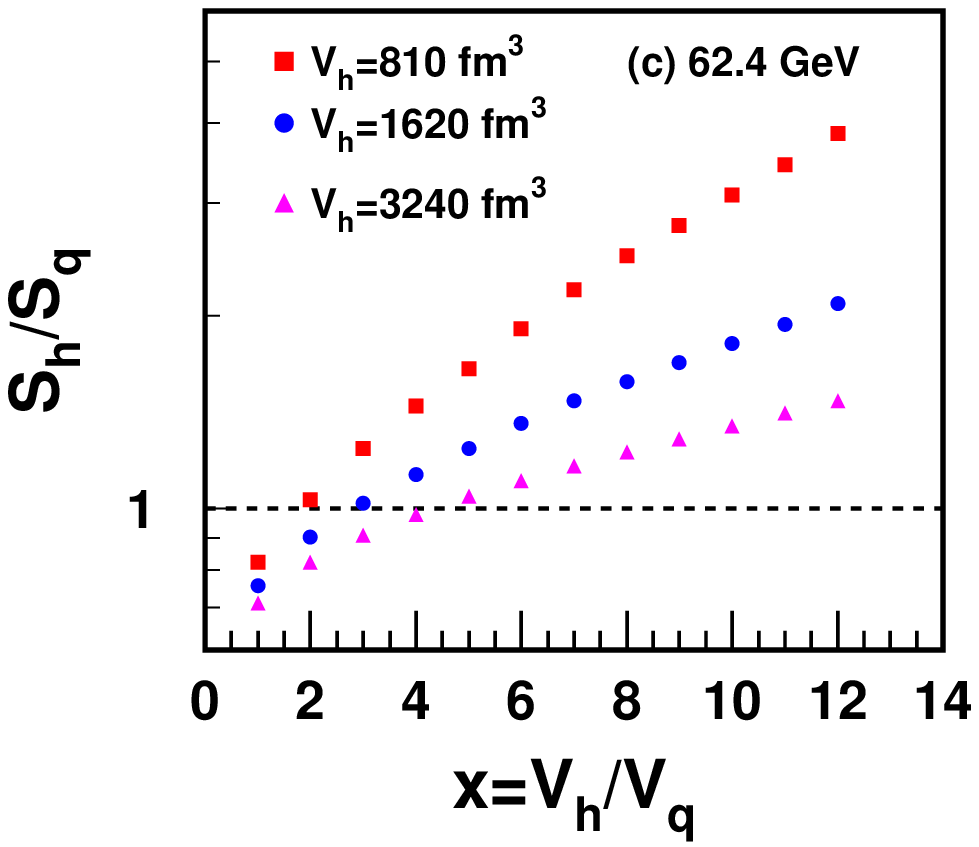}

\includegraphics[scale=0.37]{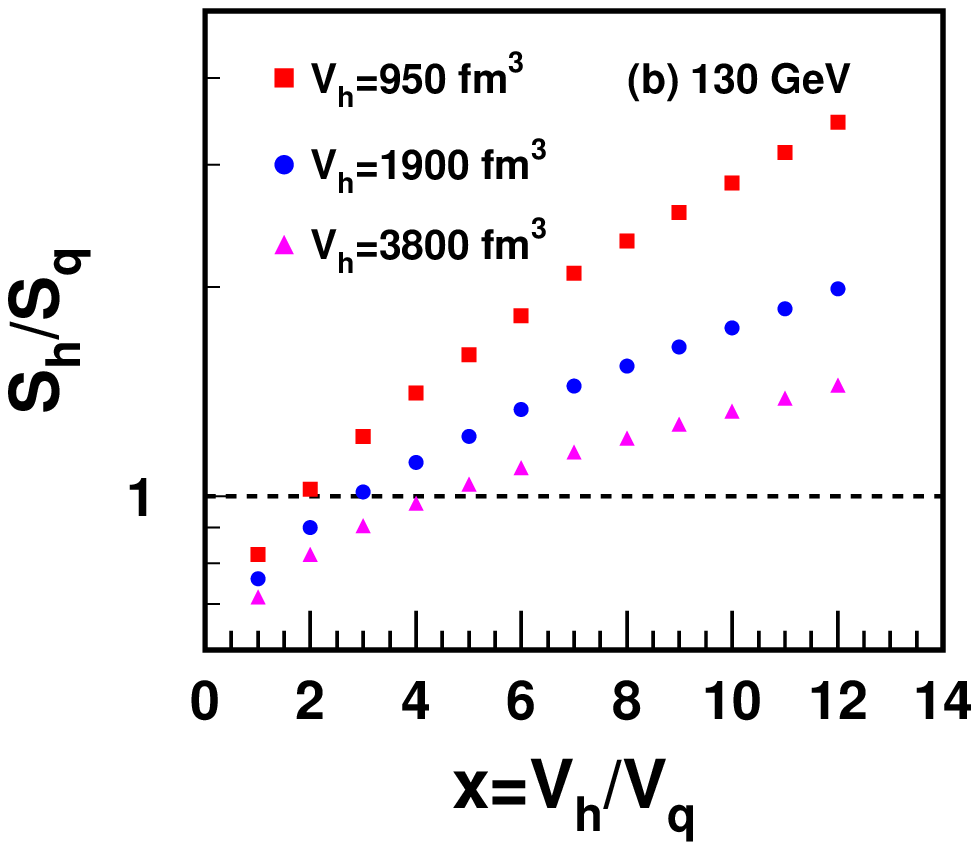}\includegraphics[scale=0.37]{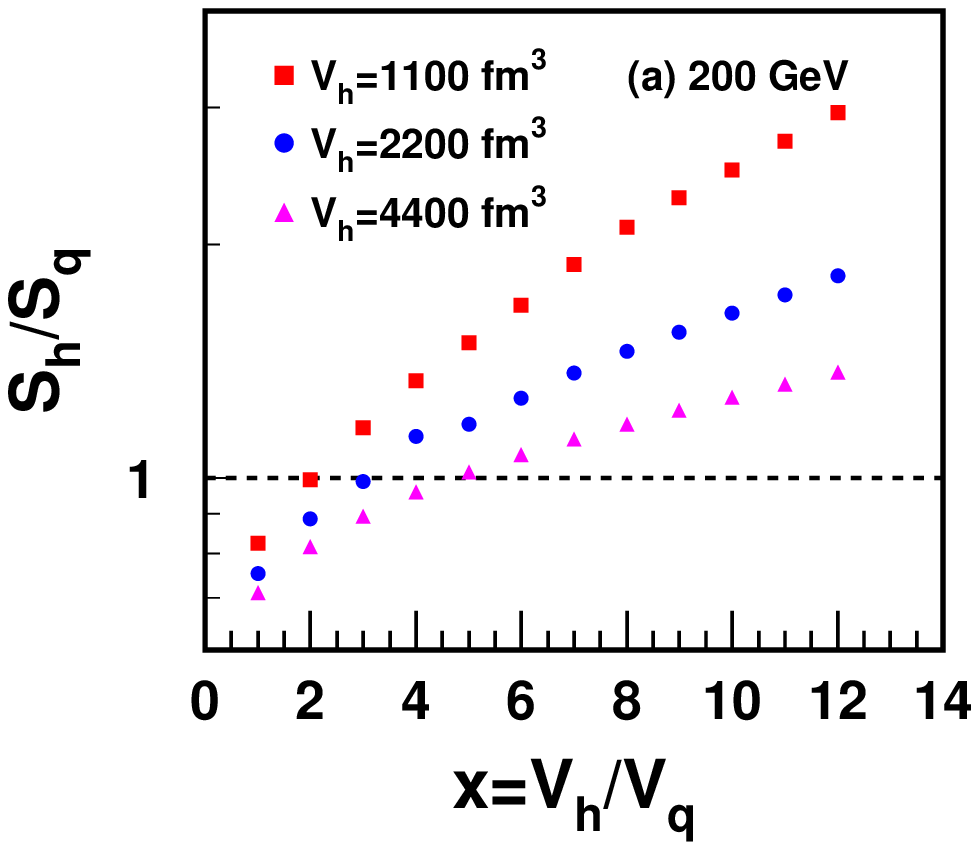}
\end{figure}

In order to calculate the entropy via Eq. (\ref{eq:fermi-dirac}-\ref{eq:classical})
from the known momentum distributions $f_{i}(\mathbf{p})$ in the
QCM, we make an approximation or give an estimate for $g_{i}(\mathbf{r},\mathbf{p})$
by $g_{i}(\mathbf{r},\mathbf{p})\approx[(2\pi)^{3}/(d_{i}V)]f_{i}(\mathbf{p})$,
where $V=V_{q},V_{h}$ are the effective volumes of quark and hadronic
phases respectively. We can then obtain the entropy ratio $R=S_{h}/S_{q}$
as functions of volume ratio $x=V_{h}/V_{q}$. The volumes $V_{h}$
of the hadronic phase is taken to be the chemical freezeout values
of the fireball which are extracted from the thermal model \cite{BraunMunzinger:2001ip,BraunMunzinger:2003zd}
at four collisional energies, $V_{h}(\mathrm{fm}^{3})=$1125 (17.3
GeV), 1620 (62.4 GeV), 1900 (130 GeV), 2200 (200 GeV). The results
are shown in Fig. \ref{fig:entropy-ratio}. For the above set of values
of $V_{h}$, only if $x>2.5-3.0$, which corresponds to about 1.35-1.44
for the ratio of the fireball radii in the hadronic to quark phase,
should the entropy of the hadronic phase be larger than that of the
quark one. 

As a contrast we also consider an ideal case for constituent quarks
whose phase space distribution follows Fermi-Dirac distribution, $g_{i}(\mathbf{r},\mathbf{p})=1/[\exp(\varepsilon_{i}/T)+1]$.
The hadron spectra can be determined by combination of quarks. Then
we use the Duhem-Gibbs relation (\ref{eq:entropy}) and entropy formula
(\ref{eq:fermi-dirac}) to obtain the entropies for quark and hadronic
phases. The result from the Duhem-Gibbs relation is $(E,PV,TS)=$(0.61,0.145,0.755)
(quark), (0.59,0.136,0.726) (hadron) for unit volume of quark phase.
We can also see that $TS$ is almost a constant, same as in the real
case. The result from the entropy formula shows the same behavior
as in Fig. \ref{fig:entropy-ratio} except that the entropy of the
hadronic phase is larger than that of the quark one happens when $x\gtrsim16$. 

A few comments about our approach and results are needed. Our results
show that the increase of the entropy requires an effective and adequate
expansion of the fireball during hadronization, which also implies
that hadronization takes finite time to complete. Actually in hydrodynamic
simulation of the evolution of the fireball, hadronization is indeed
a dynamic process with non-trivial space-time profile, i.e. it does
not take place uniformly in the same space-time but on the freezeout
hypersurface, so different part of the fireball hadronizes in different
time. Our current approach does not take such a space-time picture
but just provide an averaged effect which is more simple and transparent
than a real hydrodynamic simulation. If combination is treated locally
one can still compute the entropies before and after hadronization
for each space-time cell using Eq. (\ref{eq:entropy}) or Eqs. (\ref{eq:fermi-dirac}-\ref{eq:classical}).
The total entropy is a sum over all cells and the result is similar
to the ideal case in our paper. This implies that the entropy can
be described in a global and effective way as in our current model.
In other words the entropy is a global quantity which should be insensitive
to the local fine structure. As we have emphasized in the beginning
that we do not address in our QCM the entropy issue in the context
of phase transition. We just made a few comments about it. If local
equilirium is reached for a closed system the entropy would not change
during the transition from the quark to hadronic phase, it is the
\textit{entropy density} that changes (decreases) during the transition
acompanied by the volume expansion. If the system is not in local
equilibrium the phase transition is not well defined (it is indeed
a crossover) but still one can obtain the total entropy which should
increase. Such a study in the QCM is independent of whether the entropy
increases or decreases beyond the combination process. 

In summary, we have investigated the issue of the entropy in the framework
of the quark combination-like model. As an example for such types
of models we used the one developed by Shandong group whose exclusive
nature makes a transparent calculation feasible. We used two available
methods to calculate the entropies for the quark and hadronic phases,
one from the Duhem-Gibbs relation, another from the entropy formula
in terms of particle phase space distributions. We found that the
total entropy from the Duhem-Gibbs relation always increases in hadronization
if the average temperature of the hadronic phase is lower than that
of the quark phase. The increase of the entropy during hadronization
can also be confirmed from the entropy formula if the volume of the
hadronic phase is larger than 2.5-3.0 times that of the quark phase.
This implies that the expansion of the fireball takes place during
hadronization and it takes finite time for the quark phase to hadronize.
So whether the entropy increases or decreases during combination depends
on the temperature before and after combination and on how much expansion
the system undergoes during combination. The current study provides
an example to shed light on the entropy issue in the quark combination
model. 

Acknowledgment. We thank S. Bass, R. Hwa, C.-M. Ko and C.-B. Yang
for insightful discussions. ZTL and JS are supported in part by the
National Natural Science Foundation of China (NSFC) under the grant
10975092. YXL is supported in part by NSFC under the grants 10425521
and 10675007. FLS is supported by NSFC under the grant 10775089. QW
is supported in part by the '100 talents' project of Chinese Academy
of Sciences (CAS) and by the NSFC under the grants 10675109 and 10735040.


\begin{thebibliography}{20}
\bibitem{Anisovich:1972pq}V.~V.~Anisovich and V.~M.~Shekhter, Nucl.\ Phys.\  B {\bf 55}, 455 (1973).   

\bibitem{Bjorken:1973mh}J.~D.~Bjorken and G.~R.~Farrar, Phys.\ Rev.\  D {\bf 9}, 1449 (1974).   

\bibitem{Hwa:2002tu}R.~C.~Hwa and C.~B.~Yang, Phys.\ Rev.\  C {\bf 67}, 034902 (2003).   

\bibitem{Greco:2003xt}V.~Greco, C.~M.~Ko and P.~Levai, Phys.\ Rev.\ Lett.\  {\bf 90}, 202302 (2003).   

\bibitem{Fries:2003vb}R.~J.~Fries, B.~Muller, C.~Nonaka and S.~A.~Bass, Phys.\ Rev.\ Lett.\  {\bf 90}, 202303 (2003).

\bibitem{Greco:2003mm}V.~Greco, C.~M.~Ko and P.~Levai, Phys.\ Rev.\  C {\bf 68}, 034904 (2003).   

\bibitem{Nonaka:2005vr}C.~Nonaka, B.~Muller, S.~A.~Bass and M.~Asakawa, Phys.\ Rev.\  C {\bf 71}, 051901 (2005).   

\bibitem{Hwa:2004ng}R.~C.~Hwa and C.~B.~Yang, Phys.\ Rev.\  C {\bf 70}, 024905 (2004).   

\bibitem{Hwa:2009bh}R.~C.~Hwa, arXiv:0904.2159 [nucl-th].  

\bibitem{Xie:1988wi}Q.~B.~Xie and X.~M.~Liu, Phys.\ Rev.\  D {\bf 38}, 2169 (1988).   

\bibitem{Liang:1991ya}Z.~T.~Liang and Q.~B.~Xie, Phys.\ Rev.\  D {\bf 43}, 751 (1991).   

\bibitem{Wang:1995ch}Q.~Wang and Q.~B.~Xie, J.\ Phys.\ G {\bf 21}, 897 (1995).   

\bibitem{Wang:1996pg}Q.~Wang, Z.~G.~Si and Q.~B.~Xie, Int.\ J.\ Mod.\ Phys.\  A {\bf 11}, 5203 (1996).   

\bibitem{Shao:2004cn}F.~l.~Shao, Q.~b.~Xie and Q.~Wang, Phys.\ Rev.\  C {\bf 71}, 044903 (2005).   

\bibitem{Song:2007cx}J.~Song, F.~l.~Shao, Q.~b.~Xie, Y.~f.~Wang and D.~m.~Wei, arXiv:nucl-th/0703095.   

\bibitem{Shao:2009uk}C.~e.~Shao, J.~Song, F.~l.~Shao and Q.~b.~Xie, Phys.\ Rev.\  C {\bf 80}, 014909 (2009).   

\bibitem{Schnedermann:1993ws}E.~Schnedermann, J.~Sollfrank and U.~W.~Heinz, Phys.\ Rev.\  C {\bf 48}, 2462 (1993).   

\bibitem{Demir:2008tr}N.~Demir and S.~A.~Bass, Phys.\ Rev.\ Lett.\  {\bf 102}, 172302 (2009).   

\bibitem{BraunMunzinger:2001ip}P.~Braun-Munzinger, D.~Magestro, K.~Redlich and J.~Stachel, Phys.\ Lett.\  B {\bf 518}, 41 (2001).   

\bibitem{BraunMunzinger:2003zd}P.~Braun-Munzinger, K.~Redlich and J.~Stachel, arXiv:nucl-th/0304013. 
\end{thebibliography}
\end{document}